\documentstyle[prl,aps,graphicx,twocolumn,bm]{revtex}

\newcommand{\chem}[1]    {${\mathrm{#1}}$}
\newcommand{\unit}[1]    {~{\mathrm{#1}}}
\newcommand{\axis}[1]    {${#1}$}

\newcommand{\BSCCO}      {\chem{Bi_2Sr_2CaCu_2O_{8+{\mathit y}}}}

\newcommand{\YBCO}       {\chem{YBa_2Cu_3O_{7-\delta}}}

\newcommand{\LB}         {\linebreak[3]}

\begin{document}
\wideabs{
\title{Asymmetric Field Profile in Bose Glass Phase  
   of Irradiated {\boldmath \YBCO}:~\\
   Loss of Interlayer Coherence around 1/3 of Matching Field}

\author{K. Itaka${}^1$ \and T. Shibauchi${}^{1}$\cite{shibauchi} 
\and M. Yasugaki${}^1$ \and T. Tamegai${}^{1,2}$ \and and S. 
Okayasu${}^3$}
\address{${}^1$Department of Applied Physics, The University of Tokyo,
    7-3-1 Hongo, Bunkyo-ku, Tokyo 113-8656, Japan}
\address{${}^2$CREST, Japan Science and Technology Corporation (JST)}
\address{${}^3$JAERI, 2-4 Shirakata Shirane, Tokai-mura, 
Naka-gun, Ibaraki 319-1195, Japan}

\date{\today}
\maketitle
\begin{abstract}
Magneto-optical imaging in \LB \YBCO~ \LB with tilted columnar 
defects (CD's) shows an asymmetric critical-state field profile. 
The observed hysteretic shift of the profile ridge (trough) from 
the center of the sample is explained by in-plane magnetization 
originated from vortex alignment along CD's. The extracted ratio 
of the in-plane to out-of-plane magnetization component has a 
maximum at 1/5 of matching field ($B_\Phi$) and disappears above 
$B_\Phi/3$, suggesting a reduction of interlayer coherence well 
bellow $B_\Phi$ in the Bose glass phase. Implications are 
discussed in comparison with the vortex liquid recoupling 
observed in irradiated \BSCCO.
\end{abstract}

}
\narrowtext
%
An effective way to enhance the critical current density 
  ($J_c$) in high temperature superconductors (HTSC's) is 
  introduction of the correlated disorder such as columnar defects (CD's)
  \cite{Civale91A,Konczykowski91A,Yang96A} and
  planar defects \cite{Chong97A,Itaka99A}.
Heavy-ion irradiation is one of the most controllable techniques 
to introduce CD's in the sample whose 
  diameter is of the order of coherence length in HTSC's.
When the density of CD's, or dose-equivalent matching field $B_\Phi$, 
  is increased, the enhancement of $J_c$ persists at higher fields 
  \cite{Civale91A}.
However, the field dependence of $J_c$,
  which is proportional to the irreversible magnetization \cite{Bean64A},
  has a maximum not at $B_\Phi$, but at a field significantly 
  smaller than $B_\Phi$ 
  \cite{Civale91A,Konczykowski91A,Krusin-Elbaum96A,Chikumoto98A}.
\par
Recent studies on the vortex phase diagram of highly anisotropic  
\BSCCO (BSCCO) with CD's have suggested a nearly temperature 
independent boundary at $\sim B_\Phi/3$ 
\cite{Sato-Kosugi-Tsuchiya,Chikumoto98A}. In the vortex liquid 
(VL) regime above the irreversibility line, dramatic enhancement 
of the \axis{c}-axis coherence at $(1/5 \sim 1/3) B_\Phi$ has been 
demonstrated by Josephson plasma resonance (JPR) 
studies\cite{Sato-Kosugi-Tsuchiya}. In addition,
  the \axis{c}-axis resistivity decreases at similar fields consistent
  with the enhancement of interlayer coherence \cite{Morozov98A-99A}. 
Reduction of reversible magnetization 
  \cite{Chikumoto98A,Li96A,Beek00A} is also reported in the 
  same field range.
Theoretically, a Monte Carlo simulation
  by Sugano {\it et al.} \cite{Sugano98A} suggests a field-driven
  transition with enhancement of vortex trapping rate by CD's at $B_\Phi/3$.
Their calculation shows that the interlayer coherence
  jumps at the transition consistent with JPR experiments in the 
  VL phase of BSCCO.
In less anisotropic \YBCO (YBCO), no JPR 
  experiments have been reported because the plasma frequency in 
  YBCO is much higher than the accessible frequency range of JPR experiments
 \cite{Sato-Kosugi-Tsuchiya}.
\par
An interesting conclusion in Ref.~\onlinecite{Sugano98A} is that in the 
solid phase below the irreversibility line, which is called as 
Bose glass (BG) phase \cite{Nelson93A},
  the sign of the interlayer phase coherence change becomes opposite:
 {\it i.e.,} it {\em decreases} at $B_\Phi/3$.
In the BG phase,
  the peak in $J_c(H)$ is  actually located at $(1/5 \sim 1/3) B_\Phi$ in BSCCO
 \cite{Chikumoto98A},
 and similar behavior can be found in YBCO as well \cite{Civale91A}.
In the collective pinning theory\cite{Larkin74A}, the reduction 
of the collective pinning length related to the interlayer 
coherence causes the enhancement of the critical 
current\cite{Blatter94A}. Thus, the origin of the $J_c(H)$ peak 
in HTSC's may be related to this $B_\Phi/3$ boundary. Note that 
in unirradiated BSCCO a steep magnetization 
increase\cite{Tamegai93A} and an abrupt reduction of the phase 
coherence\cite{Shibauchi99A-Gaifullin00A} occurs simultaneously 
at the second magnetization peak field. However, so far there is 
no direct evidence for anomalous behavior of the interlayer 
coherence in the BG phase. 
\par
In this Letter,
  we provide experimental evidence for the loss of interlayer coherence
  at $(1/5 \sim 1/3) B_\Phi$ in the BG phase
  of YBCO by using magneto-optical (MO) imaging of 
  the critical state field profile.
We found that the field profile in YBCO with slightly tilted CD's is asymmetric,
  which is explained by the alignment of vortices along CD's.
The asymmetry, 
  which can be utilized as a probe for the interlayer coherence, 
  has a maximum at $B_\Phi/5$ and disappears above 
  $B_\Phi/3$. This result strongly suggests that the field-driven 
  boundary exists in YBCO in the same field range as BSCCO.
\par
\begin{figure}
\resizebox{70mm}{!}{\includegraphics[width=70mm,keepaspectratio]{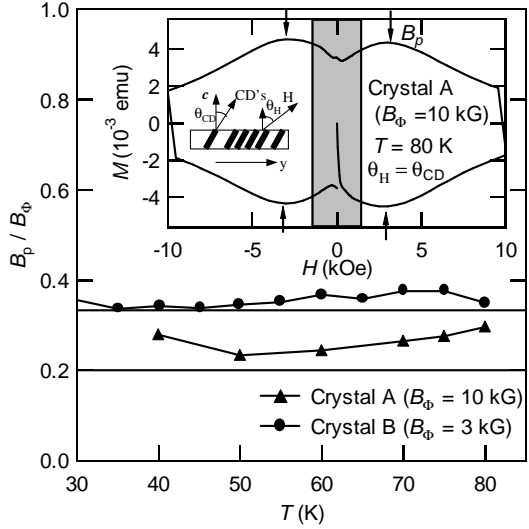}}
\caption{ Characteristic peak field $B_p$ normalized by the 
matching field $B_\Phi$ as a function of temperature (see text). 
(Inset) Magnetization hysteresis loop at $T=80\unit{K}$ in 
crystal A ($B_\Phi=10\unit{kG}$). The 
shaded area shows the field region where we performed MO imaging. 
Arrows indicate the peak field $B_p$. Schematic figure in the 
inset shows the configuration of the sample. } \label{fig1}
\end{figure}

YBCO single crystals were grown by the flux method using gold 
crucibles
  \cite{Holzberg90A}.
Rectangular twinned 
  single crystals were cut into typical dimensions of
  $1.0 \times 0.5 \times 0.015 \unit{mm^{3}}$,
  so that the edges of the samples 
  are along the $a$ and $b$ axes.
The critical temperature of the pristine samples is about 91 K.
Crystals  were irradiated with $600\unit{MeV}$ iodine ions at doses corresponding 
  to $B_\Phi=10\unit{kG}$ (crystal A) and $B_\Phi=3\unit{kG}$ (crystal B)
  using the tandem Van de Graaff accelerator with superconducting booster at JAERI.
The irradiated direction of both samples was tilted $10^\circ$ from the \axis{c}
  axis in the $y$-$c$ plane (see inset of Fig.~\ref{fig1}). 
Accordingly,
  we can distinguish the effect of CD's from that of twin boundaries (TB's).
The longitudinal magnetization $M$ parallel 
  to the applied field $H$ was measured by using a commercial 
  SQUID
  magnetometer. 
We defined 
  $\theta_{CD}$ as the angle of CD's from the \axis{c} axis,
  $\theta_{H}$ as the angle of applied field from the \axis{c} axis.
The critical state field profile was imaged by using 
  an MO indicator garnet film with in-plane magnetization placed on 
  the top surface of the sample in the field range
  $|H_{\Vert c}| \le 1.5 \unit{kOe}$,$|H_{\perp c}| \le 1.0 \unit{kOe}$\cite{Tamegai96A}.
\par
Inset of Fig.~\ref{fig1} shows magnetization hysteresis loop at 
  $T=80\unit{K}$ in crystal A with 
  $\theta_H=\theta_{CD}$.
Irreversible magnetization shows a 
  maximum at around $B_\Phi/3$ and we define $B_p$ as this peak 
  field.
In crystal B, we determined $B_p$ as 
  a field where 
$[M(\theta_H=+\theta_{CD})-M(\theta_H=-\theta_{CD})]/M(\theta_H=+\theta_{CD})$
  shows a maximum, because the enhancement of the 
  magnetization by CD's is smaller in this sample.
Main panel of Fig.~\ref{fig1} demonstrates that 
  in both crystals $B_p/B_\Phi$ is almost independent of temperature and 
  its value is around $1/3 \sim 1/5$, which is similar to that 
  reported in BSCCO\cite{Chikumoto98A}.
\par
\begin{figure}
\resizebox{70mm}{!}{\includegraphics[width=70mm,keepaspectratio]{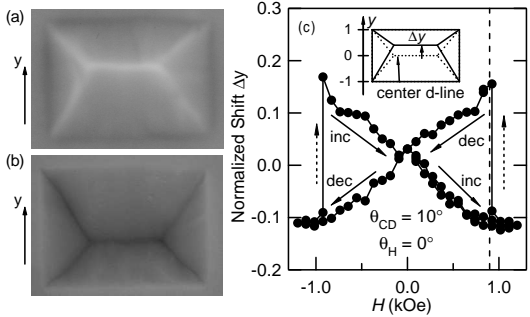}}
\caption{Typical critical-state field profiles in the (a) 
decreasing and (b) increasing field branches at $H_{\Vert c} = 900\unit{Oe}$
   in crystal A. The bright region corresponds to higher fields. The 
direction of CD's is tilted $10^\circ$ from the \axis{c} axis 
towards the \axis{y} direction.  Note the asymmetry of the 
profile.  (c) Hysteresis loop of the normalized $d$-line shift 
($\Delta y$) at $80\unit{K}$ in crystal A. Inset shows the 
definition of $\Delta y$. The sign of 
$\Delta y$ is positive when the shift direction is towards 
\axis{y} axis. Arrows with ``inc'' and ``dec'' indicate field 
sweep directions. When the field sweep direction is changed, the 
old $d$-line disappears and a new $d$-line is generated 
(dotted arrows) {\protect\cite{movie}}. 
} \label{fig2}
\end{figure}

Figures~\ref{fig2}(a) and (b) show typical critical state field profiles
  at $H_{\Vert c}=900\unit{Oe}$ in decreasing and increasing field branches,
  respectively.
The double-Y shaped current discontinuity lines ($d$-lines) are 
clearly seen, where current direction abruptly changes 
  \cite{Schuster95A,Schuster97A,Itaka99A}.
The center $d$-line is significantly shifted in the $y$ direction.
The shift of the $d$-line is much larger than $t_s \tan \theta_{CD}$,
  where $t_s$ is the thickness of the sample. The field dependence of the 
  normalized $d$-line shift $\Delta y$ in crystal A is plotted in Fig.~\ref{fig2}(c).
This curve shows hysteresis, which is symmetric with respect to $H= 0$.
When the field is increased, 
  the $d$-line at the center of the sample shifts towards one of 
  the edges (Fig.~\ref{fig2}(b)), whereas it shifts to the opposite 
  direction when the field is decreased (Fig.~\ref{fig2}(a)). The 
  direction of the shift is always along the \axis{y} axis, which is the 
  same as that of the inclination of CD's.
The hysteresis can be summarized as follows.
When $H$ and $M$ have the same sign, 
  the shift is positive, and if they are opposite,
  $\Delta y$ is negative.
When we reverse the field sweep direction, the critical 
  current direction in the sample (or the sign of $M$) is reversed, 
  and then the $d$-line shifted to one direction disappears and a 
  new $d$-line shifted to the opposite direction appears, as shown 
  by the dotted arrows in Fig.~\ref{fig2}(c) \cite{movie}.
A small positive shift $\Delta y$ at $H=0$ can be explained by 
the self field trapped in the sample. It should be noted that 
TB's cannot explain the 
  observed shift of the $d$-line, since TB's run randomly along 
  $\mathrm{[110]}$ and $\mathrm{[1\bar{1}0]}$ directions and the 
  symmetry of TB's is different from that of the $d$-line motion. 
\par
\begin{figure}
\resizebox{70mm}{!}{\includegraphics[width=70mm,keepaspectratio]{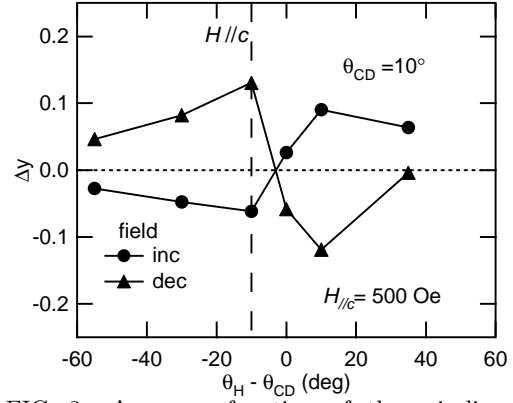}}
\caption{$\Delta y$ as a function of the misalignment angle 
($\theta_H-\theta_{CD}$) for the increasing (circles) and 
decreasing (triangles) field branches at $H_{\Vert c} = 500 \unit{Oe}$
 in crystal A. Dashed line shows the $H$ direction 
parallel to the \axis{c} axis. } \label{fig3}
\end{figure}

To clarify the relationship between $\Delta y$ and the 
 alignment of vortices along CD's, we investigated the field 
 profile under tilted fields. Figure~\ref{fig3} shows the 
 normalized $d$-line shift $\Delta y$ as a function of the 
 misalignment angle ($\theta_H-\theta_{CD}$) in crystal A.
A sign change of $\Delta y$ occurs around 
$\theta_H-\theta_{CD}\approx0$ in both increasing and decreasing 
  field branches. When $|\theta_H-\theta_{CD}|$ is large,
 $|\Delta y|$ becomes smaller.
This result shows that the misalignment of 
  the field from CD's is an important parameter to determine the 
  shift of the $d$-line.
\par
\begin{figure}
\resizebox{70mm}{!}{\includegraphics[width=70mm,keepaspectratio]{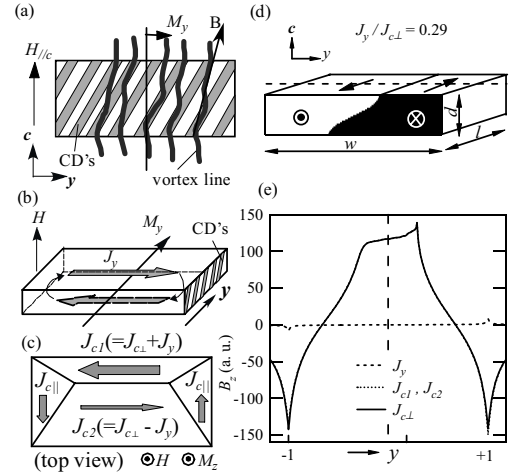}}
\caption{
 (a) In-plane magnetization $M_y$ is generated due to the misalignment of $B$ and $H$,
when the vortex is partially trapped.
In-plane (b) and out-of-plane (c) components of 
magnetization and their accompanying currents in the 
field decreasing branch at a positive field. Arrows on the sample 
show the direction and magnitude of current density. 
$J_{c1}$ and $J_{c2}$ are determined by the constraint that the 
sum of current densities in (b) and (c) cannot exceed the in-plane 
critical current density $J_{c\perp}$.
With an assumed current distribution ($J_{y}/J_{c\perp}=0.29, J \perp z,y $)
 in a finite strip sample ($w:l:d = 675: 350: 15$),  
field profile $B_z(y)$ along the broken line in (d) is calculated (e).
Induction profiles from $J_{y}$, $J_{c1}$+$J_{c2}$ and $J_{c\perp}$
are shown in broken, dotted and solid lines, respectively.
The divergence at the $d$-line is clearer in the MO image,
 because the intensity is proportional to $B_z^2$.
}
\label{fig4}
\end{figure}

Previously, an asymmetric field profile was reported in YBCO with 
  tilted CD's from the $c$ axis by Schuster {\it et al.} 
  \cite{Schuster95A}. They observed the in-plane anisotropy of $J_c$,
  and discussed it based on the difference in the nucleation 
  energy of kinks in two cases, along and across the tilted CD's\cite{note1}.
They considered that the asymmetric field profile originates from the 
  different kink nucleation between the top and bottom surfaces 
  because of the difference of the surface quality.
However, we checked that 
  $\Delta y$ at the top and bottom surfaces have opposite polarities (see Fig.~\ref{fig4}(d)),
  indicating that the asymmetry is a rather intrinsic property
  of vortex systems.
In addition,
   we performed MO imaging in the sample with non-tilted CD's
   ($\theta_{CD} \sim 0^\circ$) under tilted fields
   and confirmed 
   that the asymmetric field profile depends only on the misalignment of CD's and $H$.
\par
We propose a model to explain our observations.
In our model, the shift of the $d$-line is caused by the alignment of
  vortices along CD's, which is schematically shown in Fig.~\ref{fig4}(a).
Even when the field is applied away from the 
  direction of CD's, vortices can be aligned by CD's if the 
  misalignment is not so large. To compensate the difference
  in the directions of $H$ and $B$, in-plane magnetization ($M_y$) is 
  induced as shown in Fig.~\ref{fig4}(a), which is realized by the 
  current ($J_y$) flowing both in and across the 
\chem{CuO_2} planes (Fig.~\ref{fig4}(b)).
At the same time, 
  out-of-plane magnetization ($M_z$) is generated by the in-plane 
  current as shown in Fig.~\ref{fig4}(c).
Actual current
  density in the sample is the sum of both currents, and it is 
  limited by $J_{c\perp}$,
  where $J_{c\Vert}$ and $J_{c\perp}$ are the in-plane critical current densities
  parallel and perpendicular to the $y$ axis, respectively\cite{note1}.
The existence 
   of $J_{y}$ breaks the balance between $J_{c1}$ and $J_{c2}$, 
   because $J_{c1}$ is always anti-parallel to $J_{c2}$.
This imbalance makes the shift of the $d$-line,
   since the MO indicator can detect only the out-of-plane induction $B_z$ 
   close to the top surface.
\par
To check this idea, we calculate $B_z(y)$ assuming a current density 
distribution $J_{c\perp}(y,z)=\pm|J_{c\perp}|$ as shown in Fig.~4(d), 
where the contribution of $J_y$ is introduced. The calculated $B_z(y)$ 
in Fig.~4(e) shows a $d$-line whose position is shifted 
from the center of the superconductor, just as we observed. 
One may notice that the in-plane magnetization produces small stray fields near the edges 
($y\sim\pm1$), but in the total $B_z(y)$ this effect is negligibly small.
Our images in Fig.~2 are qualitatively consistent with this calculation.
\par
Our model naturally explains
 (1) the hysteresis observed in  Fig.~\ref{fig2}
  by the change of the direction of $J_{c\perp}$,
 (2) the opposite polarity between top and bottom surfaces, and
 (3) the $\Delta y$ sign change at $\theta_H\approx\theta_{CD}$ by the 
   change of the direction of $J_y$.
In the full critical state of a rectangular sample, 
we can estimate the relative critical current densities 
in the four regions separated by the 
double Y-shaped $d$-lines \cite{Schuster97A}. 
In Fig.~\ref{fig4}, the ratio $(1+\Delta y)/(1-\Delta y)$ 
  gives $J_{c1}/J_{c2}$, and $\Delta y$ is therefore equal to the 
  ratio of the current densities $J_y/J_{c\perp}$.
An important point is that the measurements of the shift $\Delta y$ is much easier
  than the global transverse magnetization ($M_y$) measurements  \cite{Zhukov97B} in thin samples,
  since $J_y/J_{c\perp}$ can be large due to the small thickness 
  even if the integrated $M_y$ is small.
\par
A useful implication of our model is that when vortices lose the interlayer 
coherence with zigzag-like structure along $z$ direction, the in-plane 
magnetization $M_y$ and hence $\Delta y$ will disappear.
Therefore, the asymmetry of the field profile in the critical state
 can be a powerful probe for the interlayer coherence.
\par
\begin{figure}
\resizebox{70mm}{!}{\includegraphics[width=70mm,keepaspectratio]{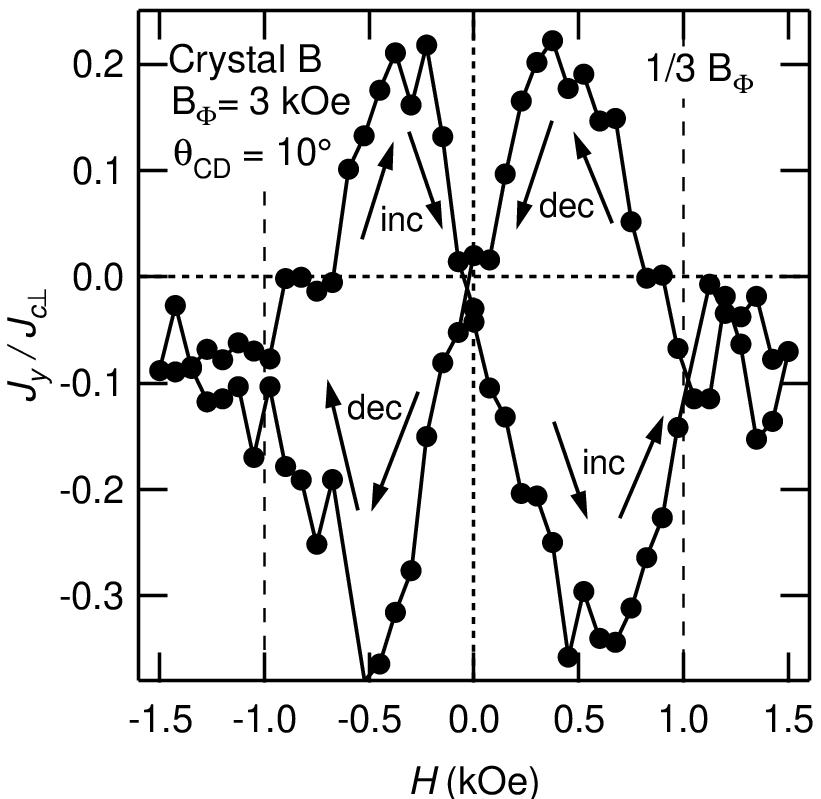}}
\caption{ $J_y/J_{c\perp}$ as a function of the applied field $H$ at 
$T=55\unit{K}$ in crystal B ($B_\Phi=3\unit{kG}$). Arrows with ``inc'' and ``dec'' indicate 
the field sweep directions. } \label{fig5}
\end{figure}

Next let us discuss what happens on the asymmetry
  when we cross $B_p$ in sample B.
Figure~\ref{fig5} shows
  $J_y/J_{c\perp}$ as a function of applied field $H$ at 
  $T=55\unit{K}$ and $\theta_{H}=0^\circ$.
The lower field part of Fig.~\ref{fig5} ($|H|<0.4$~kOe)
   is consistent with the hysteresis 
   in sample~A (Fig.~\ref{fig2}(c)). Surprisingly, the absolute value 
of $J_y/J_{c\perp}$ has a maximum at $\sim B_\Phi/5$ and becomes 
almost zero (with no hysteresis) above $B_\Phi/3$.
This behavior indicates
  that vortices are aligned along CD's below $B_\Phi/3$,
  and the interlayer coherence is suppressed above $B_\Phi/3$ \cite{note2}.
This field range of the interlayer 
  coherence anomaly is quite similar to that of field-induced 
  recoupling in the VL phase of irradiated BSCCO 
  observed by JPR experiments \cite{Sato-Kosugi-Tsuchiya}. 
Furthermore, at the same field range, the enhancement of $J_c$ is 
observed in the BG phase (Fig.~\ref{fig1}), also consistent with BSCCO \cite{Chikumoto98A}.
This similarity between YBCO and BSCCO implies that the 
  field-driven $B_\Phi/3$ anomaly does not depend on the anisotropy of the system.
Actually,
  recent JPR studies in irradiated BSCCO with different oxygen contents
  have shown no dependence of anomaly field on the anisotropy\cite{ShibauchiU1}.
Moreover,
  our results that the interlayer coherence shows
  a dramatic decrease at $\sim B_\Phi/3$, in contrast to 
  the increase in the VL state, are
  consistent with the prediction of the simulation result 
\cite{Sugano98A} in the BG phase.
\par
Finally, the reason why this anomaly field is around
  $(1/5 \sim 1/3) B_\Phi$ is still an open question.
It is well known that
  the matching effect is observed at $H = B_\Phi$\cite{note3}
  when the distribution of the pinning center is periodic\cite{Baert95A}.
In the irradiated crystals, however, CD's are 
  randomly distributed, which suggests that the statistical 
  averaging may be important to understand the underlying mechanism 
  of this number.
\par
In summary, we observed an asymmetric critical-state field 
profile in YBCO with CD's when the field is tilted away from CD's. 
The asymmetry depends on the field sweep direction and the 
misalignment of the field from CD's. We interpret this 
asymmetry in terms of the in-plane magnetization, which is 
originated from the alignment of vortices along CD's. We proposed 
that the asymmetry of the critical state field profile can be 
used as a powerful probe of the interlayer coherence. The 
coherence of vortices along CD's has a maximum at $\sim B_\Phi/5$ 
and becomes small above $B_\Phi/3$.
This result in YBCO is analogous to the results of JPR in BSCCO.
\par
We thank R. Sugano for fruitful discussions. This work is 
supported by Grant-in-Aid for Scientific Research from 
the Ministry of Education, Science, Sports and Culture, Japan.


%
%

%
%


\begin{references}
\bibitem[*]{shibauchi}{Also at 
IBM T.~J. Watson Research Center, Yorktown Heights, NY 10598, 
and MST-STC, Los Alamos National Laboratory, MS-K763, Los Alamos, NM 
87545.}

\bibitem{Civale91A}
L. Civale {\it et~al.}, Phys. Rev. Lett. {\bf 67},  648  (1991).

\bibitem{Konczykowski91A}
M. Konczykowski {\it et~al.}, Phys. Rev. B {\bf 44},  7167  (1991).

\bibitem{Yang96A}
P. Yang and C.~M. Lieber, Science {\bf 273},  1836  (1996).

\bibitem{Chong97A}
I. Chong {\it et~al.}, Science {\bf 276},  770  (1997).

\bibitem{Itaka99A}
K. Itaka {\it et~al.}, Phys. Rev. B {\bf 60},  R9951  (1999).

\bibitem{Bean64A}
C.~P. Bean, Rev. Mod. Phys. {\bf 36},  31  (1964).

\bibitem{Krusin-Elbaum96A}
L. Krusin-Elbaum {\it et~al.}, Phys. Rev. Lett. {\bf 76},  2563  (1996).

\bibitem{Chikumoto98A}
N. Chikumoto {\it et~al.}, Phys. Rev. B {\bf 57},  14507  (1998).

\bibitem{Sato-Kosugi-Tsuchiya}
M. Sato {\it et~al.}, Phys. Rev. Lett. {\bf 79},  3759  (1997);
M. Kosugi {\it et~al.}, Phys. Rev. Lett. {\bf 79}, 3763 (1997);
M. Kosugi {\it et~al.}, Phys. Rev. B {\bf 59}, 8970 (1999);
Y. Tsuchiya {\it et~al.}, Phys. Rev. B {\bf 59},  11568  (1999).

\bibitem{Morozov98A-99A}
N. Morozov {\it et~al.}, Phys. Rev. B {\bf 57}, R8146 (1998); Phys. Rev. Lett.
  {\bf 82}, 1008 (1999).

\bibitem{Li96A}
Q. Li {\it et~al.}, Phys. Rev. B {\bf 54},  R788  (1996).

\bibitem{Beek00A}
C.~J. van~der Beek {\it et~al.}, Phys. Rev. B {\bf 61},  4259  (2000).

\bibitem{Sugano98A}
R. Sugano {\it et~al.}, Phys. Rev. Lett. {\bf 80},
  2925  (1998).

\bibitem{Nelson93A}
D.~R. Nelson and V.~M. Vinokur, Phys. Rev. B {\bf 48},  13060  (1993).

\bibitem{Larkin74A}
A.~I. Larkin and Y.~N. Ovchinnikov, Sov. Phys. JETP {\bf 38},  854  (1974).

\bibitem{Blatter94A}
G. Blatter {\it et~al.}, Rev. Mod. Phys. {\bf 66},  1125  (1994).

\bibitem{Tamegai93A}
T. Tamegai {\it et~al.}, Physica C {\bf 213},  33  (1993).

\bibitem{Shibauchi99A-Gaifullin00A}
T. Shibauchi {\it et~al.}, Phys. Rev. Lett. {\bf 83}, 1010 (1999); M.~B.
  Gaifullin {\it et~al.}, Phys. Rev. Lett. {\bf 84}, 2945 (2000).

\bibitem{Holzberg90A}
F. Holtzberg {\it et~al.}, Eur. J. Solid State Inorg. Chem. {\bf 27}, 107 (1990).

\bibitem{Tamegai96A}
T. Tamegai {\it et~al.}, in {\em Critical
  Currents in Superconductors}, Proceedings of the 8th International Workshop,
  edited by T. Matsushita and K. Yamafuji (World Scientific, 1996), pp.\ 125--128.

\bibitem{Schuster95A}
T. Schuster {\it et~al.}, Phys. Rev. B {\bf 51},  16358  (1995).

\bibitem{Schuster97A}
T. Schuster {\it et~al.}, Phys. Rev. B {\bf 56},  3413  (1997).

\bibitem{movie}
A set of images are available at 
 {http://www.ap6.t.u-tokyo.ac.jp/kitaka/Research/d-line/index\_e.htm}.

\bibitem{note1}
We observed the in-plane anisotropy $J_{c \parallel}/J_{c \perp} \approx 1.2$
  in the sample B,
  consistent with the surface kink nucleations
  (see Ref.~\onlinecite{Schuster95A} and
   M.~V. Indenbom {\it et al.}, Phys. Rev. Lett. {\bf 84}, 1792 (2000).).
However, this anisotropy does not make the shift of the $d$-line
  as observed in our experiments.

\bibitem{Zhukov97B}
A.~A. Zhukov {\it et~al.}, Physica C {\bf 282--287},  2155  (1997).

\bibitem{note2}
We note that the meaning of interlayer coherence is different from
  that of trapping vortices by CD's,
  because the trapping rate can be large even if the interlayer coherence
  is very small in the case of zigzag-like vortices.

\bibitem{ShibauchiU1}
T. Shibauchi {\it et~al.}, 
  Physica C {\bf 341--348}, 973 (2000).

\bibitem{note3}
In this case, similar but reduced anomalies are also observed at $H=B_\Phi
  n~\text{and}~B_\Phi / n$ ($n$ : integer).

\bibitem{Baert95A}
M. Baert {\it et~al.}, Phys. Rev. Lett. {\bf 74},  3269  (1995).

\end{references}
\end{document}